\documentclass[12pt]{article}
\usepackage{fancyhdr}

\usepackage{mathrsfs}
\usepackage[T1]{fontenc}
\usepackage{mathpazo}
\usepackage{setspace}
\usepackage{amsfonts}
\usepackage{amssymb}
\usepackage{amsmath}
\usepackage{epsfig}
\usepackage{latexsym}
\usepackage{color}
\usepackage{graphicx}
\usepackage{nicefrac}
\usepackage[latin1]{inputenc}
\usepackage{slashed}
\usepackage{multirow}
\usepackage{comment}
\usepackage{soul}
\usepackage{hyperref}
\usepackage[nosort]{cite}
\usepackage{datetime}
\usepackage{mathtools}
\usepackage{graphicx,bbm}

\usepackage[titletoc]{appendix}

\def\hybrid{\topmargin -20pt    \oddsidemargin 0pt
        \headheight 0pt \headsep 0pt
        \textwidth 6.25in       
        \textheight 9.25in       
        \marginparwidth .875in
        \parskip 5pt plus 1pt   \jot = 1.5ex}

\hybrid

\def\baselinestretch{1.2}

\catcode`\@=11

\def\marginnote#1{}
\newcount\hour
\newcount\minute
\newtoks\amorpm
\hour=\time\divide\hour by60
\minute=\time{\multiply\hour by60 \global\advance\minute by-\hour}
\edef\standardtime{{\ifnum\hour<12 \global\amorpm={am}%
        \else\global\amorpm={pm}\advance\hour by-12 \fi
        \ifnum\hour=0 \hour=12 \fi
        \number\hour:\ifnum\minute<10 0\fi\number\minute\the\amorpm}}
\edef\militarytime{\number\hour:\ifnum\minute<10 0\fi\number\minute}

\def\draftlabel#1{{\@bsphack\if@filesw {\let\thepage\relax
   \xdef\@gtempa{\write\@auxout{\string
      \newlabel{#1}{{\@currentlabel}{\thepage}}}}}\@gtempa
   \if@nobreak \ifvmode\nobreak\fi\fi\fi\@esphack}
        \gdef\@eqnlabel{#1}}
\def\@eqnlabel{}
\def\@vacuum{}
\def\draftmarginnote#1{\marginpar{\raggedright\scriptsize\tt#1}}

\def\draft{\oddsidemargin -.5truein
        \def\@oddfoot{\sl preliminary draft \hfil
        \rm\thepage\hfil\sl\today\quad\militarytime}
        \let\@evenfoot\@oddfoot \overfullrule 3pt
        \let\label=\draftlabel
        \let\marginnote=\draftmarginnote
   \def\@eqnnum{(\theequation)\rlap{\kern\marginparsep\tt\@eqnlabel}%
\global\let\@eqnlabel\@vacuum}  }

\def\preprint{\twocolumn\sloppy\flushbottom\parindent 2em
        \leftmargini 2em\leftmarginv .5em\leftmarginvi .5em
        \oddsidemargin -.5in    \evensidemargin -.5in
        \columnsep .4in \footheight 0pt
        \textwidth 10.in        \topmargin  -.4in
        \headheight 12pt \topskip .4in
        \textheight 6.9in \footskip 0pt
        \def\@oddhead{\thepage\hfil\addtocounter{page}{1}\thepage}
        \let\@evenhead\@oddhead \def\@oddfoot{} \def\@evenfoot{} }

\def\numberbysection{\@addtoreset{equation}{section}
        \def\theequation{\thesection.\arabic{equation}}}

\def\underline#1{\relax\ifmmode\@@underline#1\else
        $\@@underline{\hbox{#1}}$\relax\fi}

\def\titlepage{\@restonecolfalse\if@twocolumn\@restonecoltrue\onecolumn
     \else \newpage \fi \thispagestyle{empty}\c@page\z@
        \def\thefootnote{\fnsymbol{footnote}} }

\def\endtitlepage{\if@restonecol\twocolumn \else \newpage \fi
        \def\thefootnote{\arabic{footnote}}
        \setcounter{footnote}{0}}  

\catcode`@=12
\relax

\def\figcap{\section*{Figure Captions\markboth
        {FIGURECAPTIONS}{FIGURECAPTIONS}}\list
        {Figure \arabic{enumi}:\hfill}{\settowidth\labelwidth{Figure
999:}
        \leftmargin\labelwidth
        \advance\leftmargin\labelsep\usecounter{enumi}}}
 \relax
\def\tablecap{\section*{Table Captions\markboth
        {TABLECAPTIONS}{TABLECAPTIONS}}\list
        {Table \arabic{enumi}:\hfill}{\settowidth\labelwidth{Table
999:}
        \leftmargin\labelwidth
        \advance\leftmargin\labelsep\usecounter{enumi}}}
 \relax
\def\reflist{\section*{References\markboth
        {REFLIST}{REFLIST}}\list
        {[\arabic{enumi}]\hfill}{\settowidth\labelwidth{[999]}
        \leftmargin\labelwidth
        \advance\leftmargin\labelsep\usecounter{enumi}}}
 \relax
\makeatletter
\newcounter{pubctr}
\def\publist{\@ifnextchar[{\@publist}{\@@publist}}
\def\@publist[#1]{\list
        {[\arabic{pubctr}]\hfill}{\settowidth\labelwidth{[999]}
        \leftmargin\labelwidth
        \advance\leftmargin\labelsep
        \@nmbrlisttrue\def\@listctr{pubctr}
        \setcounter{pubctr}{#1}\addtocounter{pubctr}{-1}}}
\def\@@publist{\list
        {[\arabic{pubctr}]\hfill}{\settowidth\labelwidth{[999]}
        \leftmargin\labelwidth
        \advance\leftmargin\labelsep
        \@nmbrlisttrue\def\@listctr{pubctr}}}
 \relax
\makeatother
\newskip\humongous \humongous=0pt plus 1000pt minus 1000pt

\newif\ifdtup

\relax

\def\be{\begin{equation}}
\def\ee{\end{equation}}
\def\ba{\begin{eqnarray}}
\def\ea{\end{eqnarray}}

\def\del{\partial}

\def\k{\kappa}

\def\a{\alpha}

\def\b{\beta}

\def\g{\gamma}
\def\G{\Gamma}
\def\d{\delta}

\def\e{\epsilon}

\def\om{\omega}

\def\l{\lambda}

\def\s{\sigma}
\def\S{\Sigma}

\def\no{\noindent}

\def\qq{\qquad}

\def\IR{\relax{\rm I\kern-.18em R}}

\def \ov {\over}

\def\diag{{\rm diag}}

\def\IR{\relax{\rm I\kern-.18em R}}
\def\IL{\relax{\rm I\kern-.18em L}}

\def\inv{^{\raise.15ex\hbox{${\scriptscriptstyle -}$}\kern-.05em 1}}

\begin{document}

\renewcommand{\theequation}{\thesection.\arabic{equation}}
\csname @addtoreset\endcsname{equation}{section}

\newcommand{\beq}{\begin{equation}}
\newcommand{\eeq}[1]{\label{#1}\end{equation}}
\newcommand{\ber}{\begin{equation}}
\newcommand{\eer}[1]{\label{#1}\end{equation}}
\newcommand{\eqn}[1]{(\ref{#1})}
\begin{titlepage}
\begin{center}


${}$
\vskip .2 in

{\large\bf String theory on the Schr\"{o}dinger pp-wave background}

\vskip 0.4in

{\bf George Georgiou}, \,  {\bf Konstantinos Sfetsos} \, and \, {\bf Dimitrios Zoakos}
\vskip 0.15in

 {\em
Department of Physics,
National and Kapodistrian University of Athens, \\
15784 Athens, Greece
}

\vskip 0.12in

{\footnotesize \texttt {ggeo@phys.uoa.gr, ksfetsos@phys.uoa.gr, zoakos@gmail.com}}


\vskip .5in
\end{center}

\centerline{\bf Abstract}

\no
We study string theory on the pp-wave geometry obtained by taking the Penrose limit around a certain null geodesic
of the non-supersymmetric Schr\"{o}dinger background.
We solve for the spectrum of bosonic excitations and find compelling agreement with the dispersion relation of the giant magnons in the  Schr\"{o}dinger background obtained previously in \cite{Georgiou:2017pvi}.
Inspired by the pp-wave spectrum we conjecture an exact in the t'Hooft coupling dispersion relation
for the magnons in the original  Schr\"{o}dinger background. 
We show that the pp-wave background admits exactly 16 Killing spinors. 
We use the explicit form of the latter in order to derive the supersymmetry algebra of the background which explicitly depends on the deformation parameter. 
Its bosonic subalgebra is of the Newton--Hooke type.

\vskip .4in
\noindent
\end{titlepage}
\vfill
\eject

\newpage

\tableofcontents

\noindent

\def\baselinestretch{1.2}
\baselineskip 20 pt
\noindent


\setcounter{equation}{0}
\section{Introduction}

The complete characterization of any conformal field theory (CFT) can be achieved by  the knowledge of  two pieces of information. The first one is the set of the theory's primary operators and their conformal dimensions. The latter can be read from the two-point correlation functions. The second piece needed is the structure constants coefficients which specify the operator product expansion (OPE) of two primary operators and which can be read from the three-point correlation functions. All higher point correlation functions can, in principle, be determined from these data. The aforementioned correlation functions are usually calculated order by order in perturbation theory as a series of one or more parameters, the couplings of the theory.  Calculating the observables of the theory at large values of the coupling constants or as an exact function of the couplings is, indeed, a rare occasion. Such an occasion is that of the maximally supersymmetric gauge theory in four dimensions,
$\mathcal {N}=4$ Super Yang-Mills (SYM). This is one of the most thoroughly studied CFTs mainly due to  its duality with 
type-IIB string theory on the $AdS_5 \times S^5$ background \cite{Maldacena:1997re}. Exploiting the key feature of integrability which  the theory possesses, an intense activity took place allowing the determination of its planar spectrum for any value of the 't Hooft coupling $\lambda$. This was achieved by the use of a variety of integrability based techniques ranging from the asymptotic Bethe ansatz \cite{Staudacher:2004tk} and the thermodynamic Bethe ansatz \cite{Ambjorn:2005wa} to the Y-system \cite{Gromov:2009tv} (for a detailed review on these techniques see \cite{Beisert:2010jr}).

On the contrary, less is known about the structure constants of the theory. The main obstacle is that for the calculation of the three-point functions the exact form of the eigenstates of the dilatation operator is also needed \cite{Georgiou:2008vk,Georgiou:2009tp,Georgiou:2011xj}.
Systematic studies of three-point correlators involving non-BPS operators  were performed in \cite{Okuyama:2004bd,Roiban:2004va,Alday:2005nd,Georgiou:2012zj} by computing the corrections arising from both the planar one-loop Feynman diagrams and the correct form of the one-loop eigenstates \cite{Georgiou:2012zj,Georgiou:2009tp}.
Alternatively, one may resort to the string theory side in order to extract information about non-protected OPE coefficients. However, this is intricate  because in the supergravity limit all non-protected operators acquire large anomalous dimensions and decouple. Nevertheless, there is a particularly useful and interesting limit in which one can extract information about  structure constants involving non-BPS operators. This is the BMN limit \cite{Berenstein:2002jq} in which one focuses on operators with large $R$-charge. These operators are dual to string states propagating in the pp-wave limit of the $AdS_5 \times S^5$
background.
Different proposals concerning the form of the cubic string Hamiltonian had been put forward  in \cite{Spradlin:2002ar,Pankiewicz:2002tg,DiVecchia:2003yp}.
The issue of how to correctly relate the string amplitudes obtained from  the pp-wave cubic Hamiltonian to the structure constants of the $\mathcal {N}=4$ SYM was settled in \cite{Dobashi:2004nm,Lee:2004cq}. This was accomplished by combining a number of results available from both  the string and the field theory sides \cite{Georgiou:2004ty,Georgiou:2003kt,Georgiou:2003aa,Chu:2002pd}.\footnote{Furthermore, using coordinate and algebraic Bethe ansatz techniques the entanglement entropy of the $\mathcal {N}=4$ SYM spin chain was studied in \cite{Georgiou:2016kge}. An exact expression for the entanglement entropy of a state with  two excitations in the BMN limit was also derived in this work.}More recently certain, non-perturbative in nature, methods for bootstrapping three-point correlators were developed in \cite{Escobedo:2010xs,Jiang:2014mja,Basso:2015zoa,Kazama:2016cfl}. In addition, by exploiting the AdS/CFT correspondence, the  strong coupling result for three-point correlators involving three {\it heavy} states in the $SU(2)$ or the  $SL(2)$ subsectors was obtained in  \cite{Kazama:2013qsa,Kazama:2011cp}. This was accomplished by calculating the area of the corresponding minimal surface through Pohlmeyer reduction. Another front where some progress has been made is the one where the three-point correlator involves two non-protected operators that are dual to classical string solutions and one light state. The strong coupling result  for this kind of three-point functions can be obtained by integrating the vertex operator of the  light state over the classical surface that describes the free propagation of the {\it heavy} state from one point on the boundary of $AdS_5$ to another \cite{Zarembo:2010rr,Costa:2010rz,Roiban:2010fe,Georgiou:2010an,Georgiou:2011qk,Bajnok:2016xxu}.

Recently, the identification of integrable deformations of the original AdS/CFT correspondence has attracted a lot of attention. One consequence of the deformation is that supersymmetry is partially or completely broken. A case where the effect of the deformation is more radical is the correspondence between  a certain Schr\"{o}dinger spacetime and its dual  null-dipole deformed conformal field theory \cite{Maldacena:2008wh,Herzog:2008wg,Adams:2008wt}. 
The theory on the gravity side \cite{Alishahiha:2003ru} is a solution of the type-IIB equations of motion and can more easily be obtained from the $AdS_5 \times S^5$ geometry through a solution generating technique known as T-s-T transformation. One starts by performing an Abelian T-duality along one of the isometries of the five-sphere  $S^5$ followed by a shift along one of the light-like directions of the $AdS_5$ boundary and then performing a second T-duality along the  coordinate of the sphere dualized initially.
The background resulting from this procedure is called $Sch_5 \times S^5$ and is non-supersymmetric.
The holographic dual field theory  is also non-supersymmetric and realizes the Schr\"{o}dinger symmetry algebra as its symmetry group. 
This field theory dual can be obtained by introducing in the $\mathcal {N}=4$ SYM Lagrangian the appropriate $\star$-product, which can be identified with the corresponding Drinfeld--Reshetikhin twist of the underlying integrable structure of the undeformed theory, that is of the $\mathcal {N}=4$ SYM 
\cite{Beisert:2005if,Ahn:2010ws,Matsumoto:2015uja,Kyono:2016jqy,vanTongeren:2015uha}. 
Consequently, the deformed theory is fully integrable and its integrability properties are inherited from the 
parent $\mathcal {N}=4$ SYM.

Compared to the original AdS/CFT scenario very few observables have been calculated in the deformed version of the correspondence. In particular, in \cite{Fuertes:2009ex} and \cite{Volovich:2009yh}  two, three and $n$-point correlation functions of scalar operators were calculated using the gravity side of the correspondence. It is important to stress that all these operators correspond to point-like strings propagating in the $Sch_5 \times S^5$ background. Extended dyonic giant magnon and spike solutions and their dispersion relations were found in
\cite{Georgiou:2017pvi}.\footnote{Giant-magnon like solutions with a different dispersion relation were studied in
\cite{Ahn:2017bio}.} Their existence is in complete agreement with the fact that the theory remains integrable. 
In the same work an exact in the coupling $\lambda$
expression for the dimensions of the gauge operators dual to the giant magnon solution was
conjectured. In the present work this is further improved in section 4 in such a way that it is in perfect agreement with the
string spectrum in the pp-wave limit.
Furthermore, in the large $J$ limit agreement was found between this expression and the one-loop anomalous dimension of BMN-like operators providing further evidence in favor of the correspondence.
On the field theory side, only the one-loop spectrum of operators belonging in a $SL(2)$ closed sub-sector has been studied \cite{Guica:2017mtd} and the authors found agreement of the one-loop anomalous dimensions of certain long operators with the string theory prediction (see also \cite{Ouyang:2017yko}).

Subsequently,  the Schr\"{o}dinger background was utilized in order to calculate, using holography,
three-point functions involving two {\it heavy} operators and a {\it light} one \cite{Georgiou:2018zkt}. The  {\it light} operator was chosen to
be one of the dilaton modes while the {\it heavy} states were either
generalizations of the giant magnon or spike solutions constructed in \cite{Georgiou:2017pvi}. These
results are the first in the literature where
the {\it heavy} states described by extended string solutions participate in three-point correlation functions.
The results of \cite{Georgiou:2018zkt} give the leading term of the correlators in the large $\lambda$ expansion
and are in complete agreement with the form of the correlator dictated by non-relativistic conformal invariance.
Finally, pulsating strings solutions in the Schr\"{o}dinger background were recently found in \cite{Dimov:2019koi}.

The aim of this work is to study string theory on the pp-wave limit of the Schr\"{o}-dinger background. 
The Penrose limit of the full geometry is taken around the null geodesic presented in \cite{Guica:2017mtd}. 
The plan of the paper is as follows.
In section \ref{section-2}, we take the Penrose limit around the aforementioned null geodesic to obtain the pp-wave geometry which after a coordinate transformation is brought to the Brinkmann form.
The corresponding mass matrix does depend explicitly
on the light cone variable.
In section \ref{section-3}, we study the spectrum of the bosonic strings  in the light-cone by deriving and solving the string equations of motion.  It so happens, that string theory on the pp-wave background is exactly solvable.
The energies of the string excitations provide the exact in the effective coupling $\lambda'=\frac{\lambda}{J^2}$ dimensions of certain operators which have large $R$-charge equal to $J$.
In section \ref{section-4}, we show that two of the eigenfrequencies of the bosonic spectrum derived in the
previous section are in complete agreement with the dispersion relation of the giant magnon solution in the
original background \cite{Georgiou:2017pvi,Georgiou:2018zkt}, that is before taking the pp-wave limit. Subsequently,
inspired by the pp-wave spectrum we conjecture an exact in the t'Hooft coupling dispersion relation
for the magnons in the original  Schr\"{o}dinger background.
Building on the supersymmetry analysis of the appendix \ref{app-SUSY}, we subsequently derive in section \ref{section-5}
the supersymmetry algebra of the background which depends explicitly on the deformation parameter $\mu$.
We provide all (anti-)commutation relations among the 17 bosonic and 16 fermionic generators.
We conclude the paper, in section \ref{section-6}.
In the appendix \ref{app-SUSY}, we focus on the supersymmetry of the pp-wave background and find
that it admits 16 Killing spinors which we explicitly write does and no more.


\section{PP-wave limit of the Schr\"{o}dinger geometry}
\label{section-2}

In this section, we review the Schr\"{o}dinger solution and take the Penrose limit around the null geodesic of
\cite{Guica:2017mtd}, to obtain the pp-wave geometry.  We will start by considering the following ten-dimensional 
$Sch_5\times S^5$ solution \cite{Maldacena:2008wh,Herzog:2008wg,Adams:2008wt}
\be
\begin{split}
\label{initial-Sch}
& ds^2  =  - \left(1\, + \, \frac{\mu^2}{Z^4} \right) dT^2   
\\
& \qq + 
\frac{1}{Z^2} \, \Big(2 \, dT \, dV + dZ^2 + dX_1^2+dX_2^2 - (X_1^2+X_2^2)  dT^2 \Big)   + ds^2_{S^5}\ ,
\end{split}
\ee
with the $S^5$ metric is written an a fiber over $CP_2$ as
\be
ds^2_{\rm S^5} = \left(d\chi+\omega\right)^2 \, + \, ds^2_{\rm \mathbb{C}P^2}\ ,
\quad
ds^2_{\rm \mathbb{C}P^2}= d\eta^2+\sin^2\eta\,
\bigl(\Sigma_1^2+\Sigma_2^2+\cos^2\eta\,\Sigma_3^2\bigr)\ ,
\ee
where the Maurer--Cartan one-forms $\Sigma_i$, $i=1,2,3$ and $\om$ explicitly given by 
\begin{eqnarray}
\begin{split}
& \Sigma_1\equiv \frac{1}{2}(\cos\psi\, d\theta - \sin\psi\sin\theta\, d\phi)\ ,
\quad
\Sigma_2\equiv \frac{1}{2}(\sin\psi\, d\theta + \cos\psi\sin\theta\, d\phi)\ ,
\\
&
\Sigma_3\equiv \frac{1}{2}(d\psi - \cos\theta\, d\phi)\ , \qq 
\quad 
\omega \equiv \sin^2\eta\, \Sigma_3\,.
\end{split}
\end{eqnarray}
 Note that $d\S_i =  \e_{ijk}\S_j \wedge \S_k$ .
The metric is supplemented with the following NS two-form
\begin{equation}
B_2 = \frac{\mu}{Z^2}\,dT \wedge \left(d\chi + \omega \right)\ ,
\end{equation}
as well as with the RR five-form
\begin{equation}
F_5 = \frac{4}{Z^5}\,dT \wedge dV \wedge  dZ \wedge  dX_1\wedge dX_2\, + \,
4 \, \cos \eta \, \sin^3 \eta \, d\eta \wedge \Sigma_1 \wedge \Sigma_2 \wedge \Sigma_3 \wedge d\chi  \, .
\end{equation}
Let us mention here that the above background is non-supersymmetric.

Now we expand around the null geodesic that is presented in \cite{Guica:2017mtd}, by considering the following ansatz for the coordinates of
the metric \eqref{initial-Sch}
\begin{eqnarray}
\begin{split}
\label{limit}
& T = \kappa \, u \, , \qquad
\chi  = \omega  \, u  + \frac{W}{L} + \frac{1}{\omega}  \frac{v}{L^2} \, , \qquad
Z = \sqrt{\frac{\kappa}{m}} \bigg(1 + \frac{\tilde Z}{L}\bigg)\ ,
\\
& V = \mu^2  m   u - \frac{\omega}{m}  \frac{W}{L}   \, , \qquad
X_i = \sqrt{\frac{\kappa}{m}} { Y_i \over L}\ ,
\qquad 
\eta = {\tilde{\eta} \over L}\ .
\end{split}
\end{eqnarray}
The parameters $\kappa$, $\omega$ and $\mu$ are constraint due to the fact that the geodesic we use is null. Explicitly,
{
\begin{equation}
\label{kmm}
\kappa^2 = \mu^2 \, m^2 + \omega^2 \, .
\end{equation}
Taking the limit $L\rightarrow \infty$, the pp-wave background takes the following form
\begin{eqnarray}
\begin{split}
& ds^2 = 2 \, d u \, d v + dW^2 + d Y_1^2+  d Y_2^2 +d {\tilde Z}^2+d {\tilde \eta}^2+
{\tilde \eta}^2 \bigl(\Sigma_1^2+\Sigma_2^2+\Sigma_3^2\bigr)
\\
&\, \, \, \quad  - \Big(\omega^2 (Y_1^2+Y_2^2)  + \mu^2 m^2 \left(Y_1^2+Y_2^2+ 4 {\tilde Z}^2\right)\Big) du^2
+2 \, \omega \left(2 {\tilde Z} dW + {\tilde \eta}^2 \, \Sigma_3 \right)du\ .
\end{split}
\end{eqnarray}
The  corresponding expressions for $H_3$ and $F_5$ are
\begin{eqnarray}
\begin{split}
& H_3 = - 2 \, \mu \, m \Big( dW \wedge d{\tilde Z} +
 {\tilde \eta}  \, d {\tilde \eta}  \wedge \Sigma_3  + {\tilde \eta}^2 \Sigma_1 \wedge \Sigma_2 \Big) \wedge  du\ ,
\\
&
F_5 =  4\, \omega \Big( dY_1 \wedge dY_2 \wedge d{\tilde Z} \wedge dW
+ {\tilde \eta}^3 d{\tilde \eta} \wedge \Sigma_1 \wedge \Sigma_2 \wedge \Sigma_3 \Big) \wedge  du \, .
\end{split}
\end{eqnarray}
Changing notation and introducing Cartesian coordinates as
\begin{equation}
\big\{{ \tilde Z} ,  {\tilde W} , Y^1, Y^2  \big\} \ \mapsto \
\big\{ y_1 , y_2 , y_3 , y_4  \big\}\ ,
 \qquad
d {\tilde \eta}^2 \, + \,  {\tilde \eta}^2 \, \sum_{i=1}^{3}  \, \Sigma_i^2 \, \equiv \, \sum_{i=5}^{8} dy_i^2\ ,
\end{equation}
enables to rewrite the pp-wave background in the following form
\begin{eqnarray}
 \begin{split}
 & ds^2= 2 \, d u \, d v
 - \Big(\omega^2 \left(y_3^2+y_4^2\right) + \mu^2 m^2 \left( y_3^2+y_4^2 + 4 y_1^2\right)\Big) du^2
\\
&
 \qq  +\sum_{i=1}^8 dy_i^2 +2 
\omega \Big(2 y_1 \, dy_2 +  y_5 \, dy_6 -  y_6 \, dy_5+  y_7 \, dy_8 -  y_8 \, dy_7 \Big)du\ ,
\end{split}
\end{eqnarray}
with the NS and the RR forms obtaining the following forms
\begin{eqnarray} 
\label{H3+F5-Brinkmann}
\begin{split}
&  H_3=2  \mu  m \Big( dy_1 \wedge dy_2 - dy_5 \wedge dy_6 - dy_7 \wedge dy_8  \Big) \wedge  du\ ,
\\
&
F_5 =  4 \omega \Big( dy_1 \wedge dy_2 \wedge dy_3 \wedge dy_4
+ dy_5 \wedge dy_6 \wedge dy_7 \wedge dy_8 \Big) \wedge  du \, .
\end{split}
\end{eqnarray}
The $B$-field whose exterior derivative reproduces the above $H_3$ reads
\begin{equation} 
\label{B-field}
B= 2  \mu  m du \wedge \Big( -y_1 \wedge dy_2 +y_5 \wedge dy_6 + y_7 \wedge dy_8  \Big)\ .
\end{equation}
Next we perform the following change of variables
\begin{equation} \label{Brinkmann-Change}
v \, \rightarrow \, v - \omega \, y_1\,  y_2\, .
\end{equation}
While $H_3$ and $F_5$ remain the same, the metric changes slightly and becomes
\begin{eqnarray}
\begin{split}
 \label{PPmetric-v2}
 & ds^2 = 2  d u   d v
 - \Big[\omega^2 \left(y_3^2+y_4^2\right) + \mu^2 m^2 \left( y_3^2+y_4^2 + 4 y_1^2\right)\Big] du^2
 + \, \sum_{i=1}^8 dy_i^2
\\
&
\qq +  2 
\omega \Big(y_1  dy_2 - y_2 dy_1 +  y_5  dy_6 -  y_6  dy_5+  y_7  dy_8 -  y_8  dy_7 \Big)du \, .
\end{split}
\end{eqnarray}
Finally, in each of the three planes $(y_1,y_2)$, $(y_5,y_6)$ and $(y_7,y_8)$,
we change variables from Cartesian to
polar coordinates. Picking  $(y_1,y_2)$ we have for the corresponding terms in the above metric that 
\begin{eqnarray}
\begin{split}
&dy_1^2 + dy_2^2 +  2  \omega \Big(y_1 dy_2 - y_2 dy_1 \Big)du
= d\rho^2 + \rho^2 \, \left(d \phi + \omega du\right)^2 - \omega^2  \rho^2    du^2
\\
& \qq\qq\qq\phantom{xxxxxxxxxxxxxxx}
= dy_1'^2 + dy_2'^2  - \omega^2 \left(dy_1'^2 + y_2'^2\right)   du^2 \ ,
\end{split}
\end{eqnarray}
where in the last step we return to Cartesian coordinates, but with an angle $\phi' \equiv \phi + \omega  u$.
Next we drop the primes in our notation.
Since there is a term containing $y_1^2$ in the first line of \eqref{PPmetric-v2}, we need to express the angle $\phi$
as a function of $\phi'$ and $u$ and consequently in terms of $y_1$ and $y_2$. 
We perform a similar change of variable in the planes  $(y_5,y_6)$ and $(y_7,y_8)$ as well.
The above variable change enables us to get rid of the $dy\,du$-terms in the metric.  
Thus, the Brinkmann form of the pp-wave metric for the Schr\"{o}dinger geometry is finally given by
\begin{equation} \label{metric-Brinkmann}
 ds^2 = 2 d u   d v + {\cal H} du^2 + \sum_{i=1}^8 dy_i^2\ ,
\end{equation}
where ${\cal H}$ is the following function of $u$ and the coordinates $y_i$'s
\begin{equation} \label{def-H}
{\cal H} \equiv - \omega^2 \sum_{i=1}^8 y_i^2 - \mu^2 m^2 \Big(
4 \left(y_1 \cos \omega u +y_2 \sin \omega u \right)^2+ y_3^2+y_4^2 \Big)  \, .
\end{equation}
The forms $H_3$ and $F_5$ remain unaffected under the change of variables from Cartesian to polar and back
to Cartesian and are still given by \eqref{H3+F5-Brinkmann}. To eliminate the cross terms $dy_i \, du$
and bring the metric in the Brinkmann form, we have inserted $u$-dependence in the coefficient of $du^2$.
Such a dependence was also observed in other kinds of deformations of the $AdS_5 \times S^5$ background, as in \cite{Mateos:2005zn,Avramis:2007wb}.


\section{Bosonic spectrum}
\label{section-3}

In this section we calculate the spectrum of the closed strings propagating in the background given by
\eqref{metric-Brinkmann} and \eqref{B-field}. We use the Brinkmann form of the background
to analyze the spectrum which will be possible to explicitly compute, 
despite the explicit dependence of the metric coefficients on the coordinate $u$.
The aim is to compare the eigenfrequencies with the dispersion relation of the giant magnon in the initial background
(see \cite{Georgiou:2017pvi,Georgiou:2018zkt}) in the common range of validity.

To comply with standard notation we will use the symbol $X^+$ instead of $u$.
The bosonic string action reads
 \begin{equation} \label{action-gen}
 S=-\frac{\sqrt{\lambda}}{2} \int \frac{d^2 \sigma}{2 \pi} \sqrt{\gamma} \, \big(  \gamma^{\alpha \beta} \partial_{\alpha}X^\mu  \partial_{\beta}X^\nu G_{\mu\nu} -\epsilon^{\alpha \beta} \partial_{\alpha}X^\mu  \partial_{\beta}X^\nu B_{\mu\nu} \big) \, ,
\end{equation}
where $\gamma^{\alpha \beta}  =  {\rm diag}(-1,1)$ and $\epsilon_{01}  = 1$. 
Note also that $\l=RL^4/\a'^2=1/\a'^2$ (we set the $S^5$ radius $R$ to unity).
Choosing the light-cone gauge $ X^+  =  \alpha' p^+ \tau$, 
the Virasoro constraints will determine $X^-$ in terms of the eight remaining physical degrees of freedom.
In addition,  it turns out that it is notationally convenient to rescale the world-sheet variables as $\tau\to \tau/(\a' p_+)$
and $\displaystyle \s\to {\s/(\a' p_+)}$. This changes the periodicity of $\s$ from $2\pi$ to $2\pi\a'p_+$.
As a result, the action governing the dynamics of the physical degrees of freedom becomes
\begin{equation} 
S = {1\ov 4\pi \a'} \int d^2 \sigma \left [\sum_{i=1}^8
\Big((\partial_{\tau}y_i)^2 - (\partial_{\sigma}y_i)^2\Big) - {\cal W}_1 + 4  \mu  m   {\cal W}_2 \right]\, ,
\end{equation}
where ${\cal W}_1$ and ${\cal W}_2$ are the following functions of the transverse
coordinates $y_i$ and $\tau$
\begin{eqnarray} \label{def-W}
\begin{split}
& {\cal W}_1 = \omega^2 \, \sum_{i=1}^8 y_i^2  +
\mu^2 m^2  \Big(4 (y_1 \cos \omega  \tau +
y_2 \sin \omega  \tau )^2 + y_3^2  + y_4^2 \Big)\ ,
\\
 & {\cal W}_2 = y_1 \, \partial_{\sigma} y_2  -
 y_5 \, \partial_{\sigma} y_6 - y_7 \, \partial_{\sigma} y_8  \, .
\end{split}
\end{eqnarray}
From this action one can derive the equations of motion for the 8 physical coordinates.


The equations of motion for both $y_3$ and $y_4$ are decoupled for the rest and
can be solved independently
\begin{eqnarray} \label{y34}
\triangle y_{3,4}  -  \left(\omega^2+\mu^2 m^2\right)  y_{3,4}  =  0\ ,\quad
\triangle \ =  - \partial_{\tau}^2  +  \partial_{\sigma}^2 \, .
\end{eqnarray}


\no
The equations of motion for the physical coordinates $y_5$ and $y_6$ are coupled in the 
following system
\begin{eqnarray} \label{y56}
\begin{split}
& \triangle y_5  -   \omega^2  y_{5}  - 
2  \mu  m\,  \partial_{\sigma} y_6  =  0\ ,
\\
& \triangle y_6   - 
 \omega^2  y_{6}  + 
2    \mu  m\,  \partial_{\sigma} y_5  =  0 \, .
\end{split}
\end{eqnarray}
The system of equations for $y_7$ and $y_8$ is identical to that for $y_5$ and $y_6$.


  The equations of motion for the physical coordinates $y_1$ and $y_2$ are coupled and depend on $\tau$.
In order to eliminate the explicit $\tau$ dependence we perform the following {\it rotation} of the coordinates $y_1$ and $y_2$
\begin{equation}
\left( \begin{array}{c}
y_1 \\
y_2
\end{array} \right)  = 
\left( \begin{array}{cc}
\cos  \omega  \tau  & -\sin  \omega \tau \\
\sin  \omega  \tau  & \cos   \omega  \tau
\end{array} \right) 
\left( \begin{array}{c}
\tilde y_1 \\
\tilde y_2
\end{array} \right) \, .
\end{equation}
The the $\tilde y_i$'s obey the system
\begin{eqnarray} \label{y12}
\begin{split}
& \triangle \tilde y_1  +  2   \omega \, \partial_{\tau} \tilde y_2  + 
2   \mu  m \, \partial_{\sigma} \tilde y_2  - 
4   \mu^2  m^2  \tilde y_1  =  0\ ,
\\
& \triangle \tilde y_2  -  2  \omega \, \partial_{\tau} \tilde y_1  - 
2   \mu  m\, \partial_{\sigma} \tilde y_1  = 0 \, .
\end{split}
\end{eqnarray}

\subsection{Solving the equations of motion}

We start with the equations of motion for $y_3$ and $y_4$.
Clearly \eqref{y34} is solved by
\begin{eqnarray}
\label{solution-y3}
\begin{split}
& y_a  =  \sqrt{\frac{\alpha'}{2}} \, \sum^{\infty}_{n=1} \, \frac{1}{\sqrt{\omega_n^{(34)}}} \,
\Bigg[ \alpha_n^a \, e^{-\frac{i}{\alpha' p^+} \, \left(\omega_n^{(34)} \tau  +  n  \sigma \right)}  + 
 \beta_n^a \, e^{-\frac{i}{\alpha' p^+}  \left(\omega_n^{(34)} \tau  -  n  \sigma \right)} \, 
 \\
 &
\qq\qq + \alpha_n^{a  \dagger} \, e^{\frac{i}{\alpha' p^+} \, \left(\omega_n^{(34)} \tau  +  n \sigma \right)}  
+   \beta_n^{a  \dagger}  \, e^{\frac{i}{\alpha' p^+} \, \left(\omega_n^{(34)} \tau -  n  \sigma \right)} \Bigg]\ ,\quad
a=3,4\ ,
\end{split}
\end{eqnarray}
which is indeed periodic in $\s$. The frequencies are
\begin{equation} 
\label{omega3}
\omega_n^{(34)} \, = \, \sqrt{\Big. n^2 \, + \, \alpha'^2 \, p^{+2} \,
\left( \omega^2 \, + \, \mu^2 \, m^2  \right) \Big.} \, .
\end{equation}
The arbitrary coefficients $\alpha_n^a$, $\beta_n^a$,
$ \alpha_n^{a  \dagger}$ and $ \beta_n^{a \dagger}$ will become operators when we quantize the system.
 Let's mention that in \eqref{solution-y3} we have not included for simplicity the zero mode contribution to the most general solution since it is irrelevant for our considerations later in the paper. We will do the same for the solutions below
for the other transverse coordinates. The
canonical momenta associated to the $y_a$'s are calculated from the Lagrangian and read
\begin{equation}
\label{mommo}
\pi_a  =  \frac{1}{2  \pi\a'} \partial_{\tau} y_a \, ,\quad a=3,4\ .
\end{equation}
Imposing the usual equal-time commutation relations
\begin{eqnarray}
\begin{split} \label{com1}
&\left[y_a(\tau,\sigma), \pi_b(\tau,\sigma')\right]  =   i \delta^{ab}  \delta(\sigma\! -\! \sigma')\ ,
\\
&
 \left[y_a(\tau,\sigma), y_b(\tau,\sigma')\right]  = \left[\pi_a(\tau,\sigma), \pi_b(\tau,\sigma')\right]   =  0\ ,
\end{split}
\end{eqnarray}
leads to 
\begin{equation} \label{creation-annihilation-op}
\left[\alpha^a_n, \alpha^{b  \dagger}_n \right]  =  \left[\beta^a_n, \beta^{b \dagger}_n \right]  =\d^{ab}  \delta_{mn} \ .
\end{equation}
Since they are appropriately normalized they have the usual interpretation as creation and annihilation operators.  Note that the inclusion of the zero mode in the solution \eqref{solution-y3} is necessary to pass from  \eqn{com1} to  \eqn{creation-annihilation-op}.

Next, we consider the light-cone Hamiltonian density, which is the conjugate momentum to the light-cone time $X^+$.
The contribution to this from the fields $y_a$ is 
\begin{equation} \label{lc-Hamiltonian-y3-v1}
H_{\rm l.c.}^a = \frac{1}{4 \pi \alpha' }  \int_0^{2  \pi  \alpha' p^+} d\sigma
\bigg[(\partial_{\tau}y_a)^2 + (\partial_{\sigma}y_a)^2 + \left( \omega^2 + \mu^2 m^2  \right) y_a^2 \bigg]\, .
\end{equation}
Substituting \eqref{solution-y3} into \eqref{lc-Hamiltonian-y3-v1} and integrating over $\sigma$ we arrive to following
expression
\begin{equation} \label{lc-Hamiltonian-y3-v2}
H_{\rm l.c.}^a \, = \, H_0^a + \frac{1}{\alpha' \, p^+} \,   \sum_{n=1}^{\infty}  \, \omega_n^{(34)}
\left(\alpha_n^{ a\dagger} \alpha_n^{a} + \beta_n^{a \dagger} \beta_n^{a}  \right)
+ \frac{1}{\alpha' \, p^+} \,   \sum_{n=1}^{\infty}  \, \omega_n^{(34)}\ ,
\end{equation}
where $H_0^a$ is the zero-mode contribution and the last term contains the zero point energy.


Consider next the system of equations for $y_5$ and $y_6$. To decouple them we introduce a new set of
coordinates which are complex conjugates to each other and given by
\begin{equation} \label{new-coord-y56}
y_{56}^{\pm}  =  y_5  \pm  i  y_6 \, .
\end{equation}
Inverting \eqref{new-coord-y56} and substituting in \eqref{y56} we obtain the following set
of decoupled equations for the new coordinates $y_{56}^{\pm}$ 
\begin{eqnarray} \label{new-y56}
 \triangle y_{56}^\pm  -  \omega^2  y_{56}^\pm  \pm
2  i   \mu  m\,  \partial_{\sigma} y_{56}^\pm =  0 \, .
\end{eqnarray}
The solution is given for
\begin{eqnarray} 
\begin{split}
\label{solution-y56p}
& y_{56}^+ =\sqrt{\a'} \sum^{\infty}_{n=1}  \Bigg[
\alpha_n^{5 \dagger}  \frac{ e^{+\frac{i}{\alpha' p^+}  \left(\omega_n^{(+)} \tau +  n  \sigma \right)} }{\sqrt{\omega_n^{(+)} }}  +
\alpha_n^{6}   \frac{e^{-\frac{i}{\alpha' p^+}  \left(\omega_n^{(+)} \tau  -  n  \sigma \right)}}{\sqrt{\omega_n^{(+)}}}
\\
&\qq\qq\qq + \beta_n^{5  \dagger}  \frac{ e^{+\frac{i}{\alpha' p^+}  \left(\omega_n^{(-)} \tau  -  n \sigma \right)} }{\sqrt{\omega_n^{(-)}}}  +
\beta_n^{6}  \frac{e^{-\frac{i}{\alpha' p^+}  \left(\omega_n^{(-)} \tau  +  n  \sigma \right)}}{\sqrt{\omega_n^{(-)}}} \Bigg] \, ,
\end{split}
\end{eqnarray}
whereas $y_{56}^-$ is given by the conjugate expression and the frequencies by
\begin{equation} \label{omega56}
\omega_n^{(\pm)}  =  \sqrt{n^2  +  \alpha'^2  p^{+2}  \omega^2  \pm 
2  n  \alpha'  p^{+} \mu m }  \  .
\end{equation}
The eight arbitrary coefficients $\alpha_n^5$, $\alpha_n^6$, $\beta_n^5$, $\beta_n^6$,
$\alpha_n^{5 \dagger}$, $\alpha_n^{6  \dagger}$, $\beta_n^{5  \dagger}$ and $\beta_n^{6  \dagger}$
will become operators as we quantize the system. The canonical momenta associated with the fields $y_5$ and $y_6$
are given by \eqref{mommo} for $a=5,6$.
Imposing the usual equal-time commutation relations the fields are promoted to operators
\begin{equation} \label{creation-annihilation-op-56}
\left[\alpha^5_n, \alpha^{5 \dagger}_n \right]  = 
\left[\alpha^6_n, \alpha^{6 \dagger}_n \right]  = 
\left[\beta^5_n, \beta^{5  \dagger}_n \right]  = 
\left[\beta^6_n, \beta^{6  \dagger}_n \right]  = \delta_{mn} \, .
\end{equation}
The light-cone Hamiltonian density, after
setting to zero all the fields except $y_5$ and $y_6$ and rescaling $\tau$ and $\sigma$, is
\begin{equation} \label{lc-Hamiltonian-y56-v1}
H_{\rm l.c.}^{56} \, = \, \frac{1}{4  \pi  \alpha' } \, \int_0^{2  \pi  \alpha'  p^+} d\sigma 
\Bigg[ \sum_{i=5}^6 \Big[(\partial_{\tau}y_i)^2 + (\partial_{\sigma}y_i)^2 + \omega^2y_i^2\Big] +
4  \mu  m   y_5 \partial_{\sigma} y_6  \Bigg]\, .
\end{equation}
Substituting the solution into \eqref{lc-Hamiltonian-y56-v1} and integrating over $\sigma$ we arrive to following
expression
\begin{eqnarray} 
\label{lc-Hamiltonian-y56-v2}
\begin{split}
& H_{\rm l.c.}^{56}  =   H_0^{56} + \frac{1}{\alpha' p^+}   \sum_{n=1}^{\infty}  \omega_n^{(+)}
\left(\alpha_n^{5 \dagger} \alpha_n^{5} +\alpha_n^{6  \dagger} \alpha_n^{6}\right)
+ \frac{1}{\alpha'  p^+}  \sum_{n=1}^{\infty} \left(\omega_n^{(+)} + \omega_n^{(-)} \right) 
\\
&\qq\qq\ + 
\frac{1}{\alpha'  p^+} \sum_{n=1}^{\infty}  \, \omega_n^{(-)}
\left( \beta_n^{5  \dagger} \beta_n^{5}  +\beta_n^{6 \dagger} \beta_n^{6} \right) \, .
\end{split}
\end{eqnarray}
where $H_0^{56}$ is the zero-mode contribution.

Although in general it is taken for granted (at least when dealing with
decoupled modes), it is not obvious a priori that the coefficient of the number operator
$\alpha_n^{p \, \dagger} \alpha_n^{p}$ (or $\beta_n^{p \, \dagger} \beta_n^{p}$)
in the light-cone Hamiltonian \eqref{lc-Hamiltonian-y56-v2} would be the corresponding frequency.
We can think of the frequencies as the exact quanta of energy that are needed to create the corresponding modes.


Finally we turn briefly to the system of equations for $y_1$ and $y_2$ by introducing the ansatz
\begin{equation} \label{Y12-ansatz}
y_{1} \, = \, \sum_{n=-\infty}^{+\infty} \alpha_n  e^{-  i  \left( \omega_n^{(12)}  \tau  + n  \sigma \right)}\ ,
\qquad
y_{2}  =  \sum_{n=-\infty}^{+\infty} \beta_n  e^{- i  \left( \omega_n^{(12)}  \tau  + n  \sigma \right)} \, ,
\end{equation}
where $\alpha_n$ and $\beta_n$ are constants.
Substituting the ansatz \eqref{Y12-ansatz} into the coupled system of equations for $y_1$ and $y_2$ in \eqref{y12},
we arrive at a homogeneous algebraic coupled system for the constants $\alpha_n$ and $\beta_n$. 
Imposing that the corresponding determinant vanishes, we arrive at the following depressed quartic equation 
for the frequencies $\omega_n^{(12)}$
\begin{equation} \label{quartic}
(\omega_n^{(12)})^4 \, + \, \gamma_2 \, (\omega_n^{(12)})^2 \, + \, \gamma_1 \, \omega_n^{(12)} \, + \gamma_0 \, = \, 0 \, ,
\end{equation}
where
\begin{equation}
\begin{split}
& \gamma_2 =  - 2  \Big[ n^2 + 2 \, \alpha'^2  p^{+2} \left(\omega^2  +  \mu^2 \, m^2\right)\Big] \  , 
\\
&\gamma_1 =  -  8  n   \mu  m  \omega  \alpha'^2  p^{+2} \ , \quad
\gamma_0 =  n^4 \, .
\end{split}
\end{equation}
It is possible to solve analytically \eqref{quartic} and determine the frequencies along the directions $y_1$ and $y_2$.
We have that
\begin{equation}  \label{omega12}
\omega_n^{(12)} = - S_1 \pm \frac{1}{2} \, \sqrt{-4 \, S_1^2 -2 \, \gamma_2 + \frac{\gamma_1}{S_1}}
\ , \quad
\omega_n^{(12)} = S_1 \pm \frac{1}{2} \, \sqrt{-4 \, S_1^2 -2 \, \gamma_2 - \frac{\gamma_1}{S_1}} \, ,
\end{equation}
where
\begin{equation}
S_1 = \frac{1}{2} \, \sqrt{- \, \frac{2}{3} \, \gamma_2 + {1 \over 3} \, \left(Q_1 + {\Delta_0 \over Q_1}\right)}\ ,
\qq
Q_1^3 = \frac{1}{2} \, \Big[ \Delta_1 \, + \sqrt{\Delta_1^2 -4 \, \Delta_0^3}\Big] \, ,
\end{equation}
with
\begin{equation}
\Delta_1 = 2 \, \gamma_2^3 +27 \, \gamma_1^2 - 72 \, \gamma_2 \, \gamma_0
\quad \& \quad
\Delta_0 =  \gamma_2^2 +12 \, \gamma_0 \, .
\end{equation}
Since the cubic term in \eqref{quartic} is absent, the sum of the frequencies in  \eqref{omega12} is zero.
In principle after determining the frequencies in \eqref{omega12} we have to follow the same path as before; analytically write
the form of the solution, impose the proper equal time commutation relations, define the standard normalized
creation and annihilation operators and finally substitute the solution into the light-cone Hamiltonian density to
{\it read} the coefficients of the number operators.

\section{PP-wave spectrum and dispersion of the giant magnon}
\label{section-4}

In this section, we will show that two of the eigenfrequencies of the bosonic spectrum derived in the
previous section is in complete agreement with the dispersion relation of the giant magnon solution in the
original background, that is before taking the pp-wave limit. 
This solution was derived
in \cite{Georgiou:2017pvi} and further studied in \cite{Georgiou:2018zkt}.

The dispersion relation of the giant magnon solution in the Schr\"{o}dinger background
reads
\begin{eqnarray} \label{dispersion}
\sqrt{E^2-\mu^2 M^2}-J \, = \, \sqrt{1+\frac{\lambda}{\pi^2} \sin^2\frac{p}{2}} =
\sqrt{1+\frac{\lambda \, n^2}{J^2}} + \cdots\ .
\end{eqnarray}
Notice that the unit under the square root can not be reproduced from
the classical string theory computation which gives the dispersion relation for $\lambda\gg 1$, but we have included it for completeness.
Furthermore, the last equality holds in the limit of small worldsheet momenta, that is
$p=\frac{2\pi  n}{J}, \,\,\,n\in Z$ and $J,\l\to \infty$, keeping  the combination $\l\ov J^2$ fixed. The dots denote subleading 
in the large $J$ expansion terms.
At first sight it seems unlikely to be able to reproduce this expression from the pp-wave limit geometry.
The reason is that the light-cone string Hamiltonian is linear in the generators $E$, $M$ and $J$,
while the left hand side of \eqref{dispersion} is nonlinear due to the presence of the square root. Indeed,
\begin{eqnarray} \label{Ham0}
H_{\rm l.c.}  \equiv  p^-  =  p_+  =  i  \frac{\partial}{\partial x^+}  =
i  \Bigg[\k  {\partial \ov \partial T}   +  \om   {\partial \ov \partial \chi} +  \mu^2  m  {\partial \ov \partial V} \Bigg] \, .
\end{eqnarray}
Taking now into account that the conserved charges in the original Schr\"{o}dinger background are given by
\begin{eqnarray} \label{charges}
E  =  i \, {\partial \ov \partial T}\, , \quad
J  =  - \, i \, {\partial \ov \partial \chi} \ , \quad
M \, = \, - \, i  \, {\partial \ov \partial V}\, ,
\end{eqnarray}
we get the following expression for the light-cone Hamiltonian in terms of the conserved charges of the original spacetime
\begin{eqnarray} \label{Ham1}
H_{\rm l.c.} =  \k  E  - \om  J  -  \mu^2  m  M\, .
\end{eqnarray}
Furthermore, the light-cone momentum is given by
\begin{eqnarray} \label{lc-moment}
 \a'  p^+ =  \a'  p_-  = -  i  \a' \frac{\partial}{\partial x^-}  = 
 -  i {\a' \ov \om L^2} {\partial \ov \partial \chi} =  {J\ov \sqrt{\l}} \, ,
\end{eqnarray}
where and in order to simplify the notation we have set $\om=1$ and we have used the relation between
the radius $L$ and the 't Hooft coupling $\l$, that is $L^2/\a'=\sqrt{\l}$.

By identifying the light-cone Hamiltonian \eqref{Ham1} with the energy of a single string
excitation given by \eqref{omega56}, i.e. $H_{\rm l.c.}=\om^\pm_n$,
we get that
\begin{eqnarray} \label{disp0}
{\k \, \sqrt{\l}  E -  \mu^2  m  \sqrt{\l} M \ov J}  - \sqrt{\l}  = 
\sqrt{1+{n^2 \l \ov J^2}  \pm  {2  n  \mu  m  \sqrt{\l} \ov J}} \ ,
\end{eqnarray}
which is still different  looking from \eqref{dispersion}. To bring \eqref{disp0} to that form we write the
energy, angular moment and particle number of the string moving on the pp-wave geometry as the corresponding
quantities of the point-like BMN particle \cite{Guica:2017mtd,Georgiou:2017pvi} plus corrections of order $\mathcal O(\l^0)$, namely
\begin{eqnarray} \label{expansion}
E  = \sqrt{\l}  \k  +  \epsilon_1 \ , \quad
J  =  \sqrt{\l}  +  j_1 \ ,\quad
M  =  \sqrt{\l}  m +  m_1 \ .
\end{eqnarray}
The next step consists of writing  the quantities $\sqrt{\l} \k$, $\sqrt{\l} $ and $\sqrt{\l} m$ in terms of $E$, $J$
and $M$ and substitute these expressions in \eqref{disp0} to obtain
\be
\label{disp1}
{E^2  -  \mu^2  M^2 \ov J}  -  J -Y_1 = 
\sqrt{1+{n^2 \l \ov J^2}  \pm  {2  n  \mu  m  \sqrt{\l} \ov J}}  \ ,
\ee
where $Y_1$ collects the terms involving $\e_1,j_1$ and $m_1$, defined as
\be
Y_1 =  {\epsilon_1  E  -  \mu^2 M  m_1 \ov J}  -  j_1 \, .
\ee
The last step is to derive the expression for $J$ in terms of $E$, $\epsilon_1$, $M$, $m_1$ and $j_1$.
This  can be done by plugging \eqref{expansion} in \eqref{kmm} to get
\begin{eqnarray} \label{inter}
\begin{split}
& E^2  -  \mu^2  M^2 -  J^2  =  Z\ ,
\\
& Z = 2 \left(\epsilon_1  \k \sqrt{\l}  - \sqrt{\l} j_1  - \mu^2  m \sqrt{\l} m_1\right)  + 
\epsilon_1^2  -  j_1^2  -  \mu^2  m_1^2 \ .
\end{split}
\end{eqnarray}
The last equation can now be solved for $J$ to give
\begin{equation} \label{Jay}
J = \sqrt{E^2  - \mu^2 M^2} \Bigg[1  -  {1\ov 2}  {Z\ov E^2 - \mu^2 M^2 }\Bigg] \, .
\end{equation}
To derive the last equation we have approximated $\sqrt{1-{Z\ov E^2 -\mu^2 M^2 }}$ by
$1-{1\ov 2}{Z\ov E^2-\mu^2 M^2 }$, since $Z$ is of ${\cal O}(\sqrt{\l})$ while
$ E^2-\mu^2 M^2$ is of ${\cal O}(\l)$.

The result for $J$ should be substituted in \eqref{disp1} to give
\begin{equation} \label{disp-fin}
\sqrt{ E^2  -  \mu^2  M^2 }  -  J  = 
\sqrt{1 +  {n^2 \l \ov J^2}  \pm {2 n  \mu  m \sqrt{\l} \ov J}}  +  {\cal W}\ ,
\end{equation}
with
\begin{equation} \label{def-W}
{\cal W} = {1 \ov 2 \, J} \, \left(\epsilon_1^2 \, - \, j_1^2 \, - \, \mu^2 \, m_1^2\right) \, + \,
\frac{Z^2}{4 \, J \, \left(Z + J^2 \right)} \, .
\end{equation}
Note that ${\cal W} $ in \eqref{disp-fin} should be ignored since it scales as ${1/J}$ and it becomes zero
in the strict $J$ infinity limit in which the giant magnon is defined.
In addition, if ones wishes to compare \eqref{disp-fin} with the dispersion relation of the giant magnon \eqref{dispersion}
in the original background one should also ignore the third term in the square root of the right hand side of \eqref{disp-fin}.
The reason is that in the large $\l$ limit the aforementioned term scales as $\sqrt{\l}$ and is thus suppressed
with respect to the previous term under the square root. In order to find this contribution in the original Schr\"{o}dinger background one should calculate the $\a'$ corrections to the dispersion relation \eqref{dispersion}

Finally, one may perform the analysis of this section for the eigenfrequency \eqn{omega3}. The result is again in agreement with the dispersion relation of the giant magnon in the original Schr\"{o}dinger background \eqref{dispersion}. We, thus, see that the BMN spectrum carries more refined information compared to the  strong coupling result of \eqref{dispersion}.

We close this section by making a speculation for the {\it exact} in $\lambda$ dispersion relation of the magnon excitations in the original Schr\"{o}dinger
background. Inspired by the form of the PP-wave spectrum it is plausible to conjecture that the exact dispersion relation corresponding to \eqref{omega3} is
%
\begin{eqnarray} \label{dispersion-1}
\sqrt{E^2-\mu^2 M^2}-J  =  \sqrt{1+\mu^2 m^2+\frac{\lambda}{\pi^2} \sin^2\frac{p}{2}} \, ,
\quad J \rightarrow \infty\ ,
\end{eqnarray}
while the one corresponding to \eqref{omega56} is
\begin{eqnarray} \label{dispersion-2}
\sqrt{E^2-\mu^2 M^2}-J =  \sqrt{1+\frac{\lambda}{\pi^2} \sin^2\frac{p}{2}+  \mu m \frac{ \sqrt{\l}}{\pi} \, \sin{p}} \, ,
\quad J \rightarrow \infty \, .
\end{eqnarray}
Indeed, for small values of $p=\frac{2\pi  n}{J}, \,\,\,n\in Z$ one gets  \eqref{omega3} and \eqref{omega56} respectively.
It would be certainly interesting to identify the corresponding field theory operators and check if their exact in $\lambda$ conformal dimensions
are given by \eqref{dispersion-1} and \eqref{dispersion-2}.
Notice that these dispersion relations can be rewritten in a form relevant for the dual null dipole CFT by using the relation $\mu = \frac{\sqrt{\lambda}}{2 \pi}  \tilde L$ (see \cite{Guica:2017mtd}), where $\tilde L$ is the parameter entering the star product that deforms the parent theory, ${\cal N}=4$ SYM. Having done this equation \eqref{dispersion-2} has a correct weak coupling expansion in integer powers of $\lambda$.


\section{Superalgebra}
\label{section-5}

In this section we present the superalgebra of the pp-wave limit of the Schr\"{o}dinger space-time. The supersymmetry algebra consists of 17 bosonic
and 16 fermionic generators (for the latter see appendix A) whose (anti-)commutation relations are presented in this section.

\subsection{Bosonic subalgebra}

We start by deriving the bosonic generators and the corresponding algebra of the type-IIB supergravity solution
in the Brinkmann forms \eqref{metric-Brinkmann}.

\no
The background is invariant under translation along the light-cone directions $x^\pm$ which are generated by
\begin{equation}
P_+\, = \, i \, \frac{\partial}{\partial x^+} \ ,\qq
P_-\, = \, i  \, \frac{\partial}{\partial x^-} \, .
\end{equation}
Furthermore, there are three $SO(2)$ groups associated with rotations in the
$(y_3,y_4)$, $(y_5,y_6)$ and $(y_7,y_8)$ planes. The corresponding generators assume the standard expressions
\be
 J_{ij}  =  -  i  \left(y_i \frac{\partial}{\partial y^j}-y_j \frac{\partial}{\partial y^i} \right) \, , \qquad (i,j)=(3,4), (5,6), (7,8)\ .
\ee
There are also 12 additional bosonic generators which can be constructed as follows.
One may readily check that the background of \eqref{metric-Brinkmann} and \eqref{H3+F5-Brinkmann} is invariant
under the infinitesimal transformations  with $\theta_i\ll 1$
\begin{eqnarray} \label{trans1}
y_i\rightarrow y_i \, + \, \theta_i \, \cos{\omega_i x^+} \ ,\qq 
x^-\rightarrow x^- \, + \, \theta_i \,\omega_i \, y_i \, \sin{\omega_i x^+}\ ,
\end{eqnarray}
where $\omega_i$ is the square root of the frequency of the corresponding coordinate in \eqref{def-H}, that is

\begin{equation}\label{freq}
\omega_3^2=\omega_4^2=\omega^2 +\mu^2 m^2, \qquad \omega_i^2=\omega^2 \, \qq  i=5,6,7,8\, .
\end{equation}
The corresponding generators read
\begin{eqnarray}\label{gen1}
M_i  =  -  i \Big( \cos \omega_i x^+ \, \frac{\partial}{\partial y^i}  + 
\omega_i y_i  \sin{\omega_i x^+  \frac{\partial}{\partial x^-}} \Big)\, , \qquad i=3,\ldots,8.
\end{eqnarray}
In a similar way one obtains the following 6 Killing vectors
\begin{eqnarray}
\label{gen2}
N_i  =  -  i  \Big( \sin{\omega_i x^+}  \frac{\partial}{\partial y^i} -
\omega_i y_i \cos{\omega_i x^+ \frac{\partial}{\partial x^-}} \Big) \, , \qquad i=3,\ldots,8.
\end{eqnarray}
which generate the following isometries of the background
\be 
\begin{split}
\label{trans2}
y_i\rightarrow y_i \, + \, \tilde{\theta_i} \, \sin{\omega_i x^+}\ , \qq
x^-\rightarrow x^- \, - \, \tilde{\theta_i} \, \omega_i \, y_i \, \cos{\omega_i x^+}\ .
 \end{split}
\ee
Thus we have in total 17 Killing vectors, listed below
\begin{equation}
\Big\{P_+ \, , \, P_- \, , \, J_{34} \, , \, J_{56} \, , \, J_{78} \, , \, M_i \, , \, N_i \Big\}
\quad \text{for} \quad i=3,\ldots,8.
\end{equation}
These Killing vectors form a bosonic subalgebra of the full algebra whose non-zero commutation relations are given below.
In particular, the bosonic subalgebra  can be viewed as three copies of the centrally extended Newton-Hooke algebra \cite{NH,Gibbons:2003rv,Grosvenor:2017dfs} in two dimensions but with all three copies sharing the same Hamiltonian $P_+$ and having the same central extension $P_-$. Each copy is of the form
\begin{eqnarray} \label{bosonic-sub}
\begin{split}
&\Big [P_+,M_i\Big ] = -  i \omega_i   N_i \, ,  \quad
\Big[P_+,N_i\Big] =  i \omega_i  M_i \, , \quad
\Big[M_i,N_j\Big]  =  -  i\omega_i \delta_{ij} \, P_- \ ,
\\
&\Big[J_{ij},M_k\Big] = i\, \left(\delta_{ik} \, M_j- \delta_{jk} \, M_i\right ) \, , \quad
\Big[J_{ij},N_k\Big] = i\, \left(\delta_{ik} \, N_j - \delta_{jk} \, N_i \right) \, ,
\end{split}
\end{eqnarray}
where $i,j,k \in \{3,4 \}$ or $i,j,k \in \{5,6 \}$ or $i,j,k \in \{7,8 \}$, with the corresponding frequencies given by \eqref{freq}.
Notice also that the above anti-commutation relations depend on the deformation parameter $\mu$ through the frequencies $\omega_3=\omega_4$.
In the next subsection we will derive the supersymmetric extension of \eqref{bosonic-sub} which is relevant for our pp-wave background.

\no
Finally, note also 
the discrete symmetry the geometry \eqref{metric-Brinkmann} and \eqref{H3+F5-Brinkmann} is invariant under a
discrete ${\mathbb Z}_2$-symmetry, interchanging $(y_5,y_6)$ and $(y_7,y_8)$.

\subsection{Commutation relations for odd generators}

To find the anticommutators between two fermionic generators of the superalgebra,
we need to calculate the expression for the vector
$V= {\bar \epsilon}_1 \Gamma^{M}  \epsilon_2  \partial_M$,
where $\epsilon_1= \epsilon_1(\psi_1)$ and $\epsilon_2= \epsilon_2(\psi_2)$ are Killing spinors,
with $\psi_1$ and $\psi_2$ being constant spinors the $\e_i$'s depend on. 
 As shown in appendix A, the pp-wave spacetime admits only
16 Killing spinors and as a result it preserves exactly half of the maximal number of supersymmetries, the rigid ones.
The Killing spinors are  independent of the $y_i$ coordinates and as a result we find that
\begin{equation}
V  =  {\bar \psi}_1  \Gamma^{-}   \psi_2  \partial_-\ .
\end{equation}
From this expression it is possible to read off the anticommutators \cite{Figueroa-OFarrill:2001hal}
\begin{equation} \label{fermionic-sub}
\Big\{ Q, Q\Big \} = -i  \Gamma^- C^{-1}   P_- \, ,
\end{equation}
where $C$ is the charge conjugation matrix \cite{Blau:2001ne,Figueroa-OFarrill:2001hal}.
Notice that only the generator $P_-$ appears in the right hand side of \eqref{fermionic-sub} in agreement with the fact that
all other bosonic generators depend on $y_i$ while the Killing spinors $\epsilon$ and as a result the right hand side of \eqref{fermionic-sub} does not.
Furthermore, $P_+$ can not either appear in the right hand side of \eqref{fermionic-sub}
because of the projection $\Gamma^+ \epsilon=0$, which the Killing spinors satisfy.
Equation \eqref{fermionic-sub} can be written in more detail as
\begin{equation} \label{fermionic-sub-gen}
\Big\{ Q_i^\a, Q_j^\b \Big \} \, = \, -i \d_{ij} \, (\Gamma^- \, C^{-1}I_2)_{\a\b} \,  P_- \, ,
\end{equation}
where the indices $i,j=1,\ldots,16$  denote the 16 independent supersymmetries while $\a,\b=1,\ldots,64$ are the spinor indices of
\begin{equation} \label{MWQs}
Q = \left( \begin{array}{c}
Q^{(1)}\\
Q^{(2)}
\end{array} \right)\, .
\end{equation}
It is known that any pp-wave geometry preserves at least half of the maximal possible supersymmetries giving rise to the so-called kinematical supercharges \cite{Cvetic:2002si}. This is precisely our case with the
16 supercharges satisfying the relation $\Gamma^+ Q=0$ being the kinematical ones. These supercharges are not corrected by string interactions and depend only on the zero modes of the string \cite{Sadri:2003pr}.

\subsection{Even-odd commutation relations}

\noindent To compute the commutator between bosonic and fermionic generators of the superalgebra,
we have to introduce the
notion of the spinorial Lie derivative $L_{\xi}$ along a Killing vector direction $\xi$
\cite{Blau:2001ne,Figueroa-OFarrill:2001hal}
\begin{equation}
L_{\xi}  = \nabla_{\xi} +  \frac{1}{4}  \nabla_{M} \xi_{N}\, \Gamma^{MN} \, .
\end{equation}
The non-vanishing spinorial Lie derivatives of the Schr\"{o}dinger Killing spinors are
\begin{equation}
L_{P_+}  \epsilon(\psi) = \epsilon \Big(-  \frac{1}{2}  \omega  \big[\Pi  +  \Pi' \big]   \sigma^2  \psi + 
\frac{i}{2} \mu m \left[\Gamma^{12}  -  \Gamma^{56}  -    \Gamma^{78}\right] \sigma^3 \psi  \Bigg)
\end{equation}
and
\begin{equation}
L_{J_{ij}}  \epsilon(\psi)  = \epsilon \Bigg(\frac{1}{2} \Gamma^{ij}  \psi  \Bigg) \ ,\quad (i,j)=(3,4),(5,6),(7,8) 
\ .
\end{equation}
Using those expressions, the mixed commutators of the superalgebra read
\begin{eqnarray} 
\label{mixed-1}
\Big[P_+, Q\Big] =-  \frac{1}{2}  \omega  \big[\Pi  +  \Pi' \big]   \sigma^2  Q + 
\frac{i}{2} \mu  m \left[\Gamma^{12}  -  \Gamma^{56}  -    \Gamma^{78}\right]  \sigma^3  Q\  ,
\end{eqnarray}
where $Q$ is the doublet of \eqref{MWQs}.
Written in the basis of complex supercharges (see \eqref{complex-spinor}) one obtains
\begin{eqnarray} 
\begin{split}
\label{mixed-1-complex}
& \Big[P_+, Q\Big]   =   \frac{1}{2}  \omega  \big[\Pi  + \Pi' \big]   Q + 
\frac{i}{2} \mu  m \left[\Gamma^{12}  -  \Gamma^{56}  -    \Gamma^{78}\right]  Q^* \ ,
 \\
& \Big[P_+, Q^*\Big]   = -  \frac{1}{2}  \omega  \big[\Pi  +  \Pi' \big]   Q^* + 
\frac{i}{2} \mu  m \left[\Gamma^{12}  -  \Gamma^{56}  -    \Gamma^{78}\right]  Q
\end{split}
\end{eqnarray}
and
\begin{equation}
\Big[J_{ij}, Q\Big]  =-  \frac{i}{2}  \Gamma_{ij}  Q \ ,\qquad
(i,j)=(3,4),(5,6),(7,8) \ ,
\end{equation}
where the generators $Q$ are again the complex Majorana--Weyl spinors. Notice that the anti-commutation relations \eqref{mixed-1-complex} depend explicitly on the deformation parameter $\mu$ as it also happened with anti-commutators of the boconic subalgebra \eqref{bosonic-sub}.

\no
To the best of our knowledge, the supersymmetry algebra of the pp-wave background that is written 
above can not be found in the literature.
Certain supersymmetric extensions of the Newton-Hooke algebra can be found in  \cite{Sakaguchi:2006pg} but these superalgebras can be obtained as contractions of the $AdS_5\times S^5$ background or its pp-wave limit and in contradistinction to our algebra they do not accommodate the massive parameter $\mu$ appearing in \eqref{mixed-1-complex}.

To conclude the analysis, one has to verify the Jacobi identities among the generators of the superalgebra.
Note that this is guaranteed by the construction of the algebra. However as an example, we present one of the Jacobi identities that is not satisfied trivially, since there is a term depending on the deformation
parameter $\mu$. That is we focus on the following Jacobi identity
\be
 \label{Jacobi-1}
\begin{split}
&\Big\{ \Big[P_+,  Q_i^\a \Big] , Q_j^\b  \Big \}  - \Big\{ \Big[Q_j^\b,  P_+\Big] ,  Q_i^\a \Big \}   + 
\Big [\Big\{ Q_i^\a,  Q_j^\b \Big \} ,  P_+\Big ]  
\\
&=-i \d_{ij} \Big( ( {\cal A} \s^2+ {\cal B} \s^3)_{\a\g}   (\G^- C^{-1} I_2)_{\g\b}+  ({\cal A}  \s^2+  {\cal B} \s^3)_{\b\g}   (\G^- C^{-1} I_2)_{\g\a}\Big)P_-\, ,
\end{split}
\ee
where $ {\cal A}=-  \frac{1}{2} \omega \big(\Pi +  \Pi' \big) $ and $ {\cal B}=\frac{i}{2}\mu  m \left(\Gamma^{12}  -  \Gamma^{56}  -    \Gamma^{78}\right)$.
In order to pass to the right hand side of the above equation we have used the fact that the last term in the first line vanishes since the anticommutator of the $Q$'s is proportional to $P_-$ which then commutes with $P_+$. 
It is, thus, enough to show that the matrix $Y=( {\cal A} \s^2+ {\cal B} \s^3)   (\G^- C^{-1} I_2)$
is antisymmetric. The 10-dimensional Gamma matrices obey 
$\{ \G^\mu,\G^\nu\}=2 \eta^{\mu\nu}$,
$\mu,\nu=0,\ldots,9$ with $\eta^{\mu\nu}=\diag(-1,1,\dots, 1)$.
We choose a realization of the 10-dimensional Gamma matrices where all of them
have purely imaginary entries and satisfy
\begin{eqnarray} \label{transpose}
(\G^0)^T=\G^0 \quad \& \quad
(\G^i)^T=-\G^i \ ,  \quad i=1,\ldots,9.
\end{eqnarray}
Furthermore, an explicit realization of the charge conjugation matrix $C$, which satisfies the relation
$C \, \G^\mu \, C^{-1} \, = \,  - \, (\G^\mu)^T$, is the following
\begin{equation}
C=\prod_1^9\G^i \, .
\end{equation}
Using the identity $\G^-C^{-1}=-C^{-1}\G^+$ and taking the transpose of the matrix $Y$ we get
\begin{eqnarray} 
\label{Yt}
\begin{split}
& Y^T= -(\G^+I_2)^T (C^{-1}I_2)^T\big( - {\cal A}^T\s^2+{\cal B}^T\s^3  \big)=-(\G^-I_2)(C^{-1}I_2)^T\big(  - {\cal A}\s^2-{\cal B}\s^3  \big)
\\
&\quad \ =-(\G^-I_2)(C^{-1}I_2)\big(   {\cal A}\s^2+{\cal B}\s^3  \big)=-(\G^-I_2)\big(   {\cal A}\s^2+{\cal B}\s^3  \big)(C^{-1}I_2)
\\
&\quad \ =-\big(   {\cal A}\s^2+{\cal B}\s^3  \big)(\G^-I_2)(C^{-1}I_2)
=-Y\, .
\end{split}
\end{eqnarray}
In the derivation we have used the explicit expression for the charge conjugation matrix $C$ given above which implies that $(C^{-1})^T=-C^{-1}$, as well as \eqref{transpose}.
In conclusion, we have shown that the Jacobi identity \eqref{Jacobi-1} does, indeed, vanish.


\section{Conclusions}
\label{section-6}

In this work we studied string theory on the pp-wave limit of the non-supersymmetric Schr\"{o}dinger background. The Penrose limit of the full geometry was taken around the null geodesic presented in \cite{Guica:2017mtd}. We casted the result 
in the standard  Brinkmann form. Subsequently, we analyzed the spectrum of the bosonic strings  in the light-cone gauge by deriving and solving exactly the string equations of motion even though along two transverse directions there is dependence on 
the light cone coordinate $u$.
The energies of the string excitations provide the exact in the effective coupling
 $\lambda'=\lambda/J^2$ dimensions of certain operators which have large $R$-charge of ${\cal O}(J)$. Most importantly, we showed that two of the eigenfrequencies of the bosonic spectrum are in complete agreement with the dispersion relation of the giant magnon solution in the original background \cite{Georgiou:2017pvi,Georgiou:2018zkt}, that is before taking the pp-wave limit. This agreement provides further evidence in favor of considering the giant magnons of \cite{Georgiou:2017pvi} as fundamental excitations of the $Sch_5\times S^5$ background. 
The agreement of the two expressions  is rather remarkable given that the left hand side of the giant magnon dispersion relation \eqref{dispersion} is non-linear while the light-cone Hamiltonian depends linearly on the generators of the Schr\"{o}dinger background before taking the Penrose limit. We showed that the BMN spectrum carries more refined information compared to the  strong coupling dispersion relation of the giant magnons.
Finally, we speculated on the exact in the t'Hooft coupling dispersion relation of the magnon in the Schr\"{o}dinger
background, corresponding to two of the eigenfrequencies of the bosonic spectrum.

We have showed that the pp-wave background admits the usual 16 Killing spinors which we explicitly wrote down. Building on these results, we derived the supersymmetry algebra of the background which depends explicitly on the deformation parameter $\mu$. We have also given all (anti-)commutation relations among the 17 bosonic and 16 fermionic generators which constitute the symmetry superalgebra.

The present work can have several extensions. It would be very interesting to calculate, from the field theory side, the anomalous dimensions of the operators that are dual to the string excitations on the pp-wave background and see if one gets agreement in the large-$J$ limit.  Furthermore, one may calculate the fermionic spectrum and also try to determine the string field theory cubic Hamiltonian. The later endeavor will certainly be more difficult than the corresponding construction in the pp-wave geometry of the $AdS_5\times S^5$ background, since the number of supersymmetries is reduced from 32 to 16. This means that the cubic vertex will be much less constrained. Nevertheless, this line of research is worth pursuing because the cubic Hamiltonian will give information for the 3-string amplitudes, as well as on the proper way of comparing them to the three-point functions  of the corresponding field theory operators. It would also be amusing to clarify how supersymmetry emerges once one restricts himself in  the subsector containing only long BMN-like operators. We know that should be the case since the original Schr\"{o}dinger background has no supersymmetries at all while its pp-wave limit possesses 16 supercharges.


\subsection*{Acknowledgments}

The work of G.G. on this project has received funding from the Hellenic Foundation for Research and Innovation
(HFRI) and the General Secretariat for Research and Technology (GSRT), under grant
agreement No 15425.



\appendix

\section{Supersymmetry of the pp-wave solution}
\label{app-SUSY}

In this appendix we will count the number of supersymmetries preserved by the pp-wave solution \eqref{metric-Brinkmann}
and calculate the expression for the Killing spinor. Our conclusion is that the pp-wave background of the non-supersymmetric
Schr\"{o}dinger spacetime preserves exactly 16 supersymmetries with the corresponding Killing spinors given by \eqref{Kill-Spin}.

The supersymmetric variations of the dilatino and the gravitino are given by the following expressions
\begin{eqnarray} 
\label{SUSY-IIB}
&& \delta \lambda =  \frac{1}{2} \slashed \partial \Phi  \epsilon - \frac{1}{24} \slashed{H}  \sigma^3  \epsilon
+ \frac{1}{2} e^{\Phi} \Big( \slashed{F}_{1} (i \sigma^2) + \frac{1}{12}\slashed{F}_3   \sigma^1 \Big)
\epsilon
\nonumber \\
&& \delta \psi_{\mu} =  D_{\mu} \epsilon - \frac{1}{8} H_{\mu \nu \rho}  \Gamma^{\nu \rho} \sigma^3 \epsilon
- \frac{e^{\Phi}}{8} \Big( \slashed{F}_1 (i \sigma^2) + {1\over 6} \slashed{F}_3 \sigma^1 +
\frac{1}{2 \times  5!} \slashed{F}_{5} (i \sigma^2) \Big)  \Gamma_{\mu} \epsilon\ ,
\end{eqnarray}
where we use the notation $\slashed{F}_n \equiv  F_{i_1\dots i_n} \Gamma^{i_1 \dots i_n}$ and
\begin{equation} \label{def-covariant}
D_{\mu} \epsilon  =   \partial_{\mu} \epsilon + \frac{1}{4} \, \omega^{a b}_{\mu}
\, \Gamma_{a b}  \epsilon \, .
\end{equation}
The Killing spinor $\epsilon$ consists of two Majorana-Weyl spinors $\epsilon_\pm$ that can be combined in
a two-component vector of the following form
\begin{equation} \label{MWspinors}
\epsilon = \left( \begin{array}{c}
\epsilon_+ \\
\epsilon_-
\end{array} \right)
\end{equation}
and satisfy the chirality condition $\Gamma_{11} \, \epsilon \, = \, \epsilon$.
Instead of working with the real two-component spinor of \eqref{MWspinors}, we can use complex spinors.
If $\epsilon_\pm$ are the two components of the real spinor in \eqref{MWspinors}, the complex spinor is
\begin{equation}
\epsilon \, = \, \epsilon_+ \, + \, i \, \epsilon_- \, .
\end{equation}
The rules to pass from one notation to the other are
\begin{equation} \label{complex-spinor}
\sigma_1  \epsilon \mapsto  i  \epsilon^* \ , \qquad
\sigma_2  \epsilon \mapsto  -   \epsilon\ ,\qq
\sigma_3  \epsilon \mapsto   \epsilon^* \  .
\end{equation}
To analyze the supersymmetry
transformations \eqref{SUSY-IIB} we define the orthonormal basis
\begin{equation} \label{frame}
e^{+} = du \, ,
\quad
e^{-}  = dv  +  \frac{1}{2}  {\cal H}   du
\ ,\qq
e^{i}  =  dy_i\ ,\qquad i=1,\dots,  8
\end{equation}
and in this way the metric becomes
\begin{equation}
\label{PPwaveMetricFrame}
ds^2 = 2 e^{+} e^{-}  + \sum_{i = 1}^{8} (e^i)^2  =  \eta_{a b} e^a  e^b \,  ,
\end{equation}
where the non-vanishing components of $\eta_{ab}$ are $\eta_{+-} = \eta_{-+} = 1$ and $\eta_{ij} = \delta_{ij}$.
The non-vanishing components of the spin connection are
\begin{equation} \label{spin-connection}
\omega_{+ i}  =  - \, \omega_{i  +} = \omega^{- i}  = - \omega^{i -}  =  \frac{1}{2} \partial_i {\cal H} \, e^{+}
\end{equation}
and the Ricci tensor has only one non-trivial component
\begin{equation}
 R_{++} = - \frac{1}{2} \, \partial^i \partial_i {\cal H}  =  8 \omega^2 + 6  \mu^2  m^2 \ .
\end{equation}
Writing the forms $H_3$ and $F_5$ in frame components, will be useful in the analysis of the supersymmetry
transformations
\begin{eqnarray}
 \label{H3F5-FrameComponents}
\begin{split}
 & H_3 = 2 \mu  m  e^{+} \wedge
 \Big( e^{1} \wedge e^{2} - e^{5} \wedge e^{6} - e^{7} \wedge e^{8} \Big) \ ,
\\
& F_5 =  4 \omega e^{+} \wedge  \Big(  e^{1} \wedge e^{2}  \wedge e^{3} \wedge e^{4}
+ e^{5} \wedge e^{6}  \wedge e^{7} \wedge e^{8} \Big) \, .
\end{split}
\end{eqnarray}
Combing the definition for the covariant derivative \eqref{def-covariant} with
the explicit expression for the spin connection \eqref{spin-connection}, it is straightforward to obtain that
\begin{equation}
D_-  =  \partial_- \, , \quad
D_+   =  \partial_+  +  \frac{1}{4} \, \sum_{i=1}^8 \partial_i {\cal H} \, \Gamma_{- i} \, , \quad
D_i  =  \partial_i \, .
\end{equation}
We define the $\Gamma^{\pm}$ matrices as
\begin{equation} \label{def-Gamma-pm}
\Gamma^{\pm} \, \equiv \, \frac{\Gamma^9 \pm \Gamma^0}{\sqrt{2}} \ ,
\end{equation}
with their algebra being
\begin{equation}
 \{   \Gamma^0 ,  \Gamma^0  \}  =   - \, \{ \Gamma^9 ,  \Gamma^9  \}  =  - \, 2 \, \mathbbm{1}
 \quad \text{with} \quad
 \{   \Gamma^0 ,  \Gamma^9  \}  =   0
\end{equation}
and this guarantees that $\big( \Gamma^{\pm} \big)^2 = 0$. 


\subsection{Dilatino equation}

The analysis will begin with the dilatino equation.
Plugging all the ingredients in the first equation of  \eqref{SUSY-IIB} and requiring that it vanishes, we obtain the
following constraint
\begin{equation} 
\label{dilatino-Projection}
\Gamma^{+} \Big( \Gamma^{12}  -  \Gamma^{56}  - \Gamma^{78} \Big) \sigma^3  \epsilon  = 0 \, .
\end{equation}
Using the algebra of the Gamma matrices it is easy to check that squaring the quantity inside the square brackets of
\eqref{dilatino-Projection} it is impossible to get the identity matrix.
The vanishing of the dilatino equation is only guaranteed with the projection
\begin{equation} \label{Gm-Projection}
\Gamma^{+}  \epsilon  =  0\ .
\end{equation}
Hence, the pp-wave of the Schr\"{o}dinger background preserves 16 supercharges.


\subsection{Gravitino equation}

To determine the precise form of the Killing spinor we turn to the gravitino equation in \eqref{SUSY-IIB}. We
analyze it for each one of the frame components defined in \eqref{frame}.


\subsubsection{The $\delta \psi_{-}$ component}

\noindent The three-form $H_3$ does not extend on the $e^{-}$ direction, while the product of the five-form
$F_5$ with $\Gamma_{-}$ vanishes, $\slashed{F}_{5} \, \Gamma_{-} = \slashed{F}_{5} \, \Gamma^{+} = 0$.
The only remaining contribution to the gravitino equation in the $e^{-}$ direction is
\begin{equation} \label{gravitino-plus}
\delta \psi_- =  D_{-} \, \epsilon  =  \partial_{-}  \epsilon  =  \partial_{v}  \epsilon  =  0\ .
\end{equation}
We conclude that the Killing spinor can only depend on the variables $u$ and $y_i$,
i.e. $\epsilon = \epsilon (u, y_i)$.


\subsubsection{The $\delta \psi_{i}$ for $i=1, \ldots , 8$ components}

\noindent We begin by writing explicitly the gravitino variation for each one of the eight transverse components. Starting with
the variation along the direction $y_1$ we have
\begin{equation} \label{gravitino1-v1}
\partial_{y_1} \epsilon + \frac{1}{2}  \mu m  \Gamma^{- 2}  \sigma^3  \epsilon 
-  \frac{i}{4} \omega  \Gamma^{+} \Big(\Pi  +  \Pi' \Big)  \Gamma_1 \sigma^2  \epsilon  =  0\ ,
\end{equation}
where we have defined
\begin{equation} \label{Pi-definition}
\Pi  \equiv  \Gamma^{1234}
\ ,\qq
\Pi'  \equiv  \Gamma^{5678}  \, .
\end{equation}
Using the algebra of the Gamma matrices and the chirality condition it is possible to rewrite the last term of
\eqref{gravitino1-v1} as follows
\begin{equation} \label{gravitino1-v2}
- \frac{i}{4}  \omega \Gamma^{+} \Big(\Pi  +  \Pi' \Big)  \Gamma_1 \sigma^2 \epsilon  = 
 \frac{i}{2}  \omega  \Gamma^{+1}  \Pi\,  \sigma^2  \epsilon\, .
\end{equation}
Combining \eqref{gravitino1-v1} and \eqref{gravitino1-v2}, we rewrite the equation that comes from the
gravitino variation along the direction $y_1$ as follows
\begin{equation} \label{gravitino1-v3}
\partial_{y_1} \epsilon  +  \frac{1}{2}  \mu  m  \Gamma^{+ 2}  \sigma^3  \epsilon 
+   \frac{i}{2}  \omega  \Gamma^{+1}  \Pi\,  \sigma^2  \epsilon  =  0 \, .
\end{equation}
The equation from the gravitino variation along the direction $y_2$ is
\begin{equation}  \label{gravitino2}
\partial_{y_2} \epsilon  - \frac{1}{2}  \mu  m  \Gamma^{+ 1} \sigma^3  \epsilon 
+  \frac{i}{2}  \omega  \Gamma^{+2} \Pi \, \sigma^2  \epsilon  =  0 \, .
\end{equation}
The equations from the gravitino variation along the directions $y_3$ and $y_4$ are
\begin{equation} \label{gravitino34}
\partial_{y_3} \epsilon 
+ \frac{i}{2} \omega  \Gamma^{+3}  \Pi \, \sigma^2  \epsilon   = 0\ ,\qq
\partial_{y_4} \epsilon 
+  \frac{i}{2}  \omega \Gamma^{+4}  \Pi \, \sigma^2  \epsilon  = 0 \, .
\end{equation}
The equation from the gravitino variation along the direction $y_5$ is
\begin{equation}  \label{gravitino5}
\partial_{y_5} \epsilon  -  \frac{1}{2}  \mu m  \Gamma^{+ 6}  \sigma^3  \epsilon 
+  \frac{i}{2} \omega \Gamma^{+5}  \Pi' \, \sigma^2 \epsilon =  0 \, .
\end{equation}
The equation from the gravitino variation along the direction $y_6$ is
\begin{equation}  \label{gravitino6}
\partial_{y_6} \epsilon  +  \frac{1}{2} \mu  m \Gamma^{+ 5} \sigma^3\epsilon 
+   \frac{i}{2}  \omega \Gamma^{+6} \Pi' \, \sigma^2 \epsilon  =  0 \, .
\end{equation}
The equation from the gravitino variation along the direction $y_7$ is
\begin{equation}  \label{gravitino7}
\partial_{y_7} \epsilon  -  \frac{1}{2}  \mu  m  \Gamma^{+ 8} \sigma^3  \epsilon 
+  \frac{i}{2} \omega  \Gamma^{+7}  \Pi' \sigma^2 \, \epsilon = 0 \, .
\end{equation}
The equation from the gravitino variation along the direction $y_8$ is
\begin{equation}  \label{gravitino8}
\partial_{y_8} \epsilon  + \frac{1}{2} \mu  m  \Gamma^{+ 7}  \sigma^3  \epsilon 
+ \frac{i}{2} \omega \Gamma^{+8}  \Pi' \, \sigma^2  \epsilon =  0 \, .
\end{equation}
Since each one of the $\partial_{y_i} \epsilon$ is proportional to the matrix $\Gamma^+$,
acting with one more derivative $\partial_{y_j}$ on the equations
\eqref{gravitino1-v3} to  \eqref{gravitino8} we obtain
\begin{equation}
\partial_{y_i} \partial_{y_j} \epsilon   =  0\ ,\qq
 i , j = 1 , \ldots , 8 \,  .
\end{equation}
Then the Killing spinor is a linear function of the coordinates $y_i$
\begin{equation} \label{Killing-Spinor-v1}
\epsilon  =  \Big( \mathbb{1} +   \Gamma^+  {\cal R} \Big)  \chi (u)\ ,
\end{equation}
where we have defined
\begin{eqnarray}
\begin{split}
& {\cal R} =  \frac{1}{2} \mu  m  \Big(\Gamma^1  y_2 - \Gamma^2  y_1 -
\Gamma^5  y_6 + \Gamma^6 y_5 - \Gamma^7  y_8  + \Gamma^8  y_7\Big) \sigma^3
 \\
& \qq\qq -  \frac{i}{2} \omega  \Big( \sum_{I=1}^4  \Gamma^I y_I  \Pi  + 
\sum_{J=5}^8  \Gamma^J  y_J \Pi' \Big)    \sigma^2 \, .
\end{split}
\end{eqnarray}
Note that, since $\G^+\e=0$, we get from \eqn{Killing-Spinor-v1} that $\G^+\chi=0$ as well. In addition, 
since $\G^+$ anticommutes with ${\cal R}$ we obtain that $\e=\chi$.


\subsubsection{The $\delta \psi_+$ component}

\noindent The last step of the analysis is the gravitino variation along the $e^{+}$ direction.
Since the Killing spinor does not depend on $v$ \eqref{gravitino-plus}, the covariant derivative $D_+$ becomes
\begin{equation}
D_+  =   \partial_u  - \frac{1}{2} \Gamma^{+}  {\cal T}\ ,
\end{equation}
where we have defined
\begin{eqnarray}
 \label{def-T}
 \begin{split}
& {\cal T} = \omega^2 \sum_{I=1}^8  \Gamma^I y_I + \mu^2 m^2
\Big( 2  \Gamma^1  y_1+ 2  \Gamma^2  y_2 + \Gamma^3  y_3 + \Gamma^4  y_4 
\\
&
\qq\qq +  2  \cos 2 \omega  u \big(\Gamma^1 y_1 - \Gamma^2 y_2\big)
+ 2 \sin 2  \omega  u \big(\Gamma^2 y_1 + \Gamma^1 y_2 \big)\Big) \, .
\end{split}
\end{eqnarray}
The equation from the gravitino variation along the direction $e^{+}$  (recall that $\e=\chi(u)$) is\footnote{We 
have used the condition
\begin{equation}
\Gamma^{-+} \epsilon  =  \left(\mathbb{1}+ \Pi \Pi' \right) \epsilon\ ,
\end{equation}
coming from the chirality condition and \eqref{def-Gamma-pm}.}
\begin{equation}  
\label{spinor-diff-equation}
\partial_u \chi  -  \frac{1}{2} \mu  m \big(\Gamma^{12} - \Gamma^{56}  -   \Gamma^{78}\big)  \sigma^3  \chi 
-  \frac{i}{2} \omega  \big(\Pi  + \Pi' \big)  \sigma^2  \chi = 0 \ ,
\end{equation}
where we have used that $\G^+$ and ${\cal T}$ anticommute so that $D_+\chi=\del_u\chi$.
Analyzing the Killing spinor in a doublet of Majorana-Weyl spinors \eqref{MWspinors},
namely $\chi_1$ and $\chi_2$, we arrive to a coupled system of first order differential equations. 
Decoupling it we obtain that
\begin{equation}
\del_u^2 \chi_1  = {\cal I} \chi_1\ ,
\ee
where the matrix ${\cal I}$ is given by
\begin{equation} 
\label{def-Imatrix}
{\cal I} =  \frac{1}{4}
\Big(\mu^2 \,m^2 \, \big(\Gamma^{12} - \Gamma^{56} -  \Gamma^{78} \big)^2  - 
\omega^2  \big(\Pi +  \Pi' \big)^2 \Big)\ .
\ee
It can be shown that this matrix has only negative eigenvalues which we denote by $-\l_i^2$. The most general solution for $\chi_1$ is 
\begin{equation}\label{Kill-Spin}
\chi_1=  \sum_{i=1}^{32}  A_i  \, \exp \Big( i \, \l_i  \left(u  +  \varphi_i\right) \Big) \chi_0^i\ .
\end{equation}
where $\chi_0^i$ are the corresponding eigenvectors of the matrix ${\cal I}$. The sum is on the 32
eigenvalues/eigenvectors of the matrix ${\cal I}$. The other component $\chi_2$ is obtained then from 
\eqn{spinor-diff-equation} and has a similar form.
An important restriction is that one should only choose
those eigenvectors that satisfy the projection
 \eqref{Gm-Projection}, namely that $\Gamma^{+}   \chi_0^i  =  0$ which means that effectively the sum over $i$ in \eqref{Kill-Spin} runs from 1 to 16.



\end{document}